\documentclass{styles}

\usepackage[utf8]{inputenc}
\usepackage[T1]{fontenc}
\usepackage{graphicx}

\usepackage{amsmath}
\usepackage{textcomp}
\usepackage{amstext}
\usepackage{amssymb}
\usepackage{tikz}
\usepackage{lettrine}
\usepackage[export]{adjustbox}
\usepackage{ragged2e}
\usepackage{subfig}
\usepackage{float}
\usepackage{booktabs}
\usepackage{makecell, multirow}
\usepackage{natbib}
\usepackage{hyperref}

\usepackage{url}

\begin{document}

\title[Alternative Loss Function in Evaluation of Transformer Models\,\ldots]{Alternative Loss Function in Evaluation of Transformer Models}

\author[Michańków et al.]{Jakub Michańków, Paweł Sakowski, Robert Ślepaczuk}

\email{jakub.michankow@triplesun.net, p.sakowski@uw.edu.pl, rslepaczuk@wne.uw.edu.pl}

\institution{Triple Sun, University of Warsaw / Dep. of Quantitative Finance and Machine Learning / QFRG,University of Warsaw / Dep. of Quantitative Finance and Machine Learning / QFRG}

\city{Krakow,Warsaw,Warsaw}

\country{Poland,Poland, Poland}

\maketitle

\vspace{-20pt}
\begin{abstract}
The proper design and architecture of testing machine learning models, especially in their application to quantitative finance problems, is crucial. The most important aspect of this process is selecting an adequate loss function for training, validation, estimation purposes, and hyperparameter tuning. Therefore, in this research, through empirical experiments on equity and cryptocurrency assets, we apply the Mean Absolute Directional Loss (MADL) function, which is more adequate for optimizing forecast-generating models used in algorithmic investment strategies. The MADL function results are compared between Transformer and LSTM models, and we show that in almost every case, Transformer results are significantly better than those obtained with LSTM.\par\vspace*{2mm}
\textbf{Keywords:} Deep Learning, Neural Networks, LSTM, Algorithmic Investment Strategies, Loss Function
\end{abstract}

\section{Introduction}
\label{Introduction}

This research focuses on several key issues at the intersection of machine learning and quantitative finance. First, there is a theoretical focus on determining the most suitable architecture for testing machine learning forecasting models. Second, it includes practical efforts to use these forecasts to generate signals for algorithmic investment strategies. Third, it involves testing and comparing Transformer models with LSTM models to evaluate their effectiveness in investment strategies. Finally, there is practical testing of empirical data from stock and cryptocurrency markets across multiple assets.

The main goal of this research is to apply the transformer model to time series forecasting, using a recently introduced loss function (MADL). We also compare the transformer with the LSTM using two types of asset classes. There are two opposing sides in the scientific community: one saying that transformers can be successfully applied to time series forecasting, and one that they can't and shouldn't. Both sides provide significant examples and research to prove their point. We intend to engage in this discourse and conduct our comprehensive research.

Transformer models with attention mechanism were first proposed in 
\cite{vaswani2023attentionneed}. Since then, they gained traction as one of the pillars of Large Language Models (LLM). They were also at the core of tools such as ChatGTP which are considered groundbreaking in terms of AI. Similarly to LSTM and other RNNs, they were designed for working with sequential data, specifically text and language tasks.

The methodology is based on the application of two alternative models (Transformer and LSTM) to generate long/short signals for two types of assets: cryptocurrencies (Bitcoin, Ethereum, and Litecoin) and equities (JP Morgan, S\&P500 and Exxon Mobil Corp) with daily data. To keep the out-of-sample period as long as possible, a walk-forward procedure was applied.  The performance of the trading strategies is evaluated using risk-adjusted returns, drawdown metrics, and equity lines.

We contribute to the literature in the following ways. We present the application of an adequate loss function (MADL) in ML models to generate trading signals. Additionally, we verify the advantages of using the transformer model over the LSTM in algorithmic trading. Furthermore, we apply a strict methodology for six assets, controlling the overfitting effects, applying a walk-forward procedure, and extending the out-of-sample period for 9+ years for equity and 8+ years for cryptocurrency assets.

The structure of this paper was planned as follows. First, we present a short literature review. Then, methodology and data is discussed. Next, we present outcomes of our experiments on equities and cryptocurrencies. Finally, we summarize our findings in  conclusions.

\section{Literature review}
\label{Literature review}

The transformer model was first introduced by \cite{vaswani2023attentionneed} revolutionizing sequence modeling with its self-attention mechanism, which enabled better handling of long-range dependencies without relying on recurrent structures. Since then, researchers have explored its application across various domains, including time series and financial forecasting. A few years after the model’s introduction, \cite{zeng2022transformerseffectivetimeseries} critically assessed Transformer-based models for time series and suggested that simple, one-layer linear models (LTSF-Linear) might outperform Transformers in certain settings, challenging the notion that complex models always yield better results. However, \cite{wen2023transformerstimeseriessurvey} provide a comprehensive review of recent advancements in adapting Transformers for time series, highlighting modifications that improve its applicability and performance in this field.

The literature also explores the integration of attention mechanisms with other models. In a study by \cite{10.1371/journal.pone.0227222}, attention is successfully applied to recurrent models like LSTM and GRU, allowing them to capture relevant features over time, while \cite{Zhou_2020} show that LSTM with attention can outperform traditional ARIMA models. \cite{WANG2022118128} utilize the Transformer framework to predict the stock market index. Through the encoder-decoder architecture and the multi-head attention mechanism, Transformer can better characterize the underlying rules of stock market dynamics. We implement several back-testing experiments on the main stock market indices worldwide, including CSI 300, S\&P 500, Hang Seng Index, and Nikkei 225. All these experiments demonstrate that Transformer outperforms other classic methods significantly and can gain excess earnings for investors. \cite{zeng2023financialtimeseriesforecasting} propose to harness the power of CNNs and Transformers to model both short-term and long-term dependencies within a time series, and forecast if the price would go up, down, or remain the same (flat) in the future. They demonstrated the success of the proposed method in comparison to commonly adopted statistical and deep learning methods for forecasting intraday stock price change of S\&P 500 constituents. Finally, \cite{RePEc:eee:finana:v:90:y:2023:i:c:s1057521923003927} propose a novel Transformer model for financial forecasting, suggesting that self-attention mechanisms can better capture time-series information related to returns and volatility, providing more economic insights and predictability than nonlinear models like LSTM. 

This literature suggests that, while Transformers offer promising potential in time series forecasting, particularly in financial applications, practical experimentation remains limited. Consequently, further empirical studies are needed to establish their advantages over traditional neural networks like LSTM in real-world financial contexts.

Our methodology avoids critical flaws in studies on algorithmic investment strategies. It is worth pointing out that most of them do not employ proper testing structures, undermining the validity and robustness of their results. Common issues include over-optimization of models, use of inappropriate optimization criteria or loss functions, and limited or non-existent out-of-sample testing, which restricts generalizability (\cite{lopez2013look}, 
\cite{bailey2016probability}, 
\cite{dipersioArtificialNeuralNetworks2016}, 
\cite{yangDeepLearningStock2019},
\cite{wiecki2016all}, 
\cite{lopez2013look}, 
\cite{bailey2016probability}, 
\cite{raudys2016portfolio}). 
Other frequent problems involve reliance on a single instrument, forward-looking bias (\cite{chan2013algorithmic}, \cite{chan2021quantitative}, \cite{jansen2020machine}), 
absence of sensitivity analysis (\cite{dipersioArtificialNeuralNetworks2016}, \cite{zhang_multi_2018}, and \cite{yangDeepLearningStock2019}),
data snooping bias (\cite{bailey2016backtest}, \cite{chan2013algorithmic}), 
survivorship bias (\cite{chan2021quantitative}) 
and improper performance metrics
(\cite{CHAKOLE2021113761}, \cite{GROBYS2020101396}).

Addressing these issues requires careful model testing, with particular focus on appropriate hyperparameter tuning and loss function selection to improve the robustness of results.

\section{Methodology and Data}
\label{Methodology and Data}

\subsection{Methodology}
\label{methodology}

\subsection*{Transformer}
\label{transformer}

The Transformer architecture, introduced in \cite{vaswani2023attentionneed}, relies on self-attention to assign varying importance to different parts of the input sequence, enabling efficient modeling of long-range dependencies. Its parallelizable structure allows for fast training on large datasets and has played a significant role in recent progress. While it achieves top performance in NLP, its impact extends to other domains as well.

\begin{figure}[htbp]
  \centering
   \includegraphics[width=0.5\linewidth]{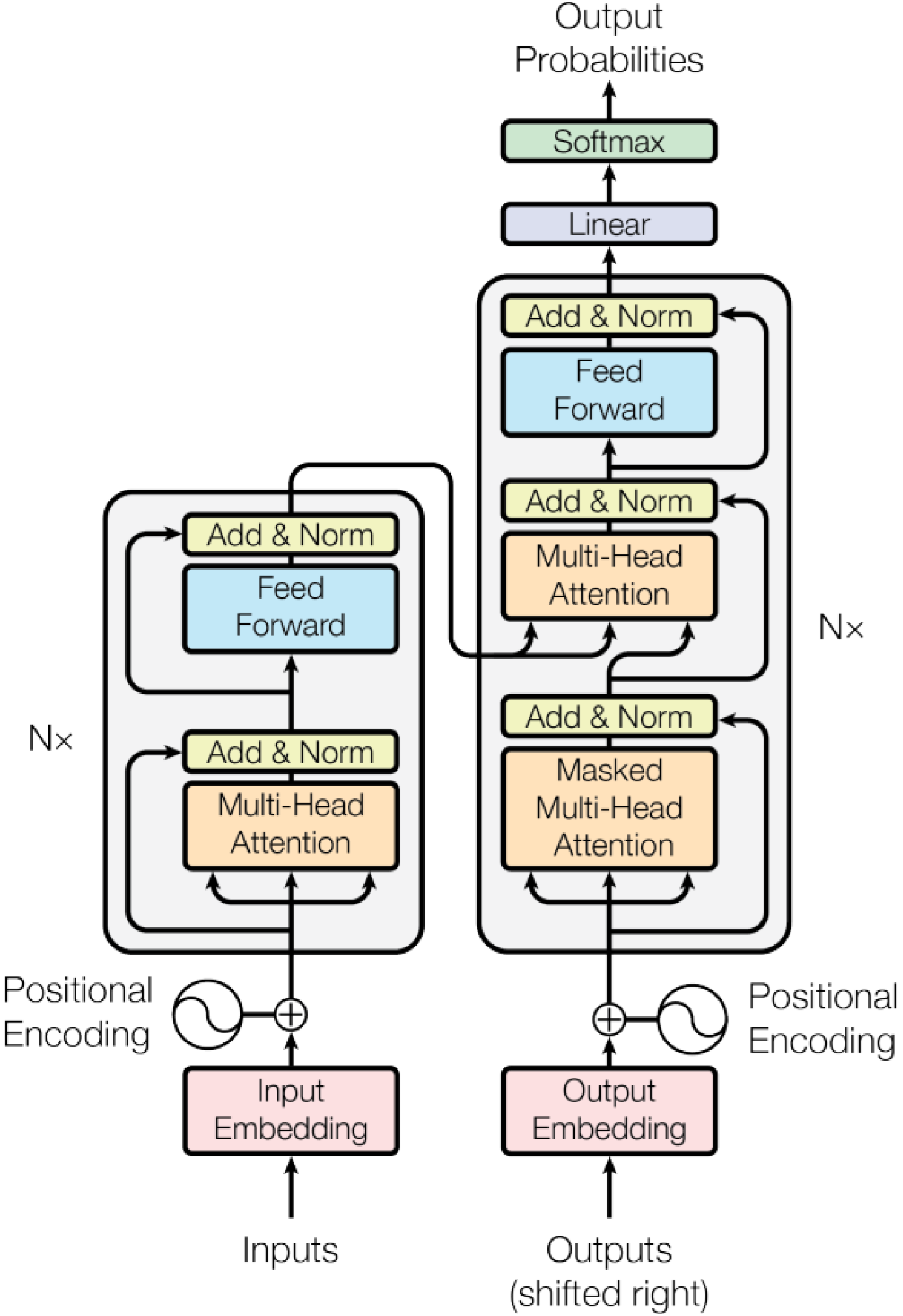}
   \caption{The structure of the Transformer model with special attention to input and output layers. Source: \cite{vaswani2023attentionneed}.}
   \label{fig:Transformer}
\end{figure}

A Transformer model consists of two main components: the encoder and the decoder. The encoder extracts features from the input, while the decoder generates output based on this representation (Fig. \ref{fig:Transformer}).

In machine translation, the encoder processes the source language, and the decoder produces the target language. In time-series forecasting, the encoder is often unnecessary, as the task involves predicting future values from past observations. The decoder’s self-attention mechanism captures temporal dependencies effectively through its autoregressive structure.

\begin{figure}[htbp]
  \centering
   \includegraphics[width=0.9\linewidth]{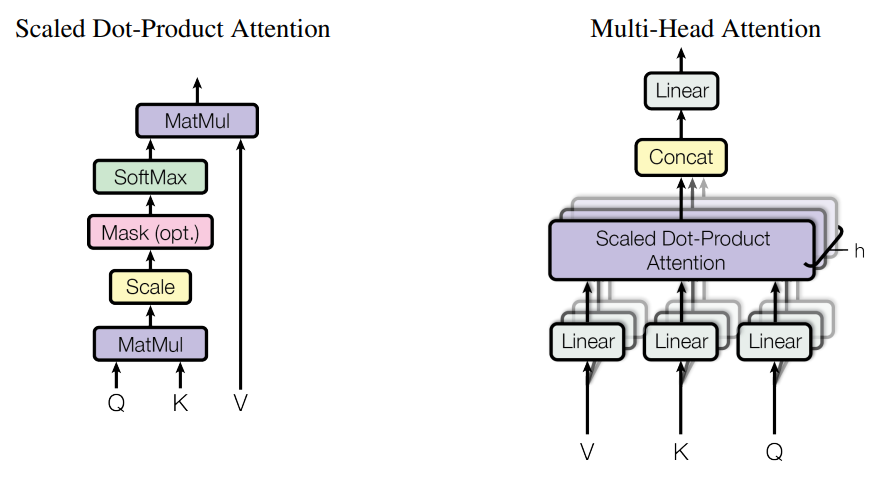}
   \caption{Transformer model with two different attention mechanisms: Scaled Dot-Product Attention and Multi-Head Attention. Source: \cite{vaswani2023attentionneed}.}
   \label{fig:AttentionMechanisms}
\end{figure}

Variations in Transformer models often arise from different attention mechanisms (see Fig. \ref{fig:AttentionMechanisms}), described as follows:

\textbf{Scaled Dot-Product Attention:}
    Inputs are queries (Q) and keys (K) of dimension $d_k$, and values (V) of dimension $d_v$. The matrix of outputs is computed as:
    
    \begin{equation}
        \text{Attention}(Q,K,V) = \text{softmax}(\frac{QK^T}{\sqrt{d_k}})V
          \label{eq:Attention}
    \end{equation}

\textbf{Multi-Head Attention:}

    \begin{equation}
        \text{MultiHead}(Q,K,V) = \text{Concat}(head_1,...,head_2)W^O,
          \label{eq:MultiHead}
    \end{equation}
    \begin{equation}
        head_i = \text{Attention}(QW^Q_i,KW^K_i,VW^V_i)
          \label{eq:Head}
    \end{equation}
where $W$ are parameter matrices and $W^Q_i \in \mathbb{R}^{d_{model}\times d_k}$, $W^K_i \in \mathbb{R}^{d_{model}\times d_k}$, $W^V_i \in \mathbb{R}^{d_{model}\times d_v}$, $W^O \in \mathbb{R}^{hd_{v}\times d_{model}} $, $d$ stands for dimension and $h$ is the number of parallel attention layers, called heads. Other abbreviations in (\ref{eq:Attention}), (\ref{eq:MultiHead}), and (\ref{eq:Head}) stand for:

    \textbf{Query (Q)}: The query represents the element of the input sequence for which attention weights are calculated—it defines what the model is focusing on. It is typically a linearly transformed version of the input, and each attention head uses separate learnable parameters to compute its own query representation.

\textbf{Key (K)}: The key is another transformed representation of the input, used to assess the relevance of each input element with respect to the query. Like the query, it is derived via a linear transformation, and each head uses independent parameters.

\textbf{Value (V)}: The value contains the information to be aggregated, based on the attention scores computed from the query and key. It is generally a transformed version of the original input sequence.

In time series forecasting, the key holds the historical context up to time step $t$, with length defined by the sequence length hyperparameter. The query corresponds to the time step $t+1$, i.e., the future value the model aims to predict. The value includes all historical data points from $t-n$ to $t$, where $n$ is the sequence length.

The multi-head attention mechanism computes attention scores between the query (representing $t+1$) and the keys (historical data up to $t$). These scores indicate how relevant each past time step is for predicting $t+1$. The values (from $t-n$ to $t$) are then weighted accordingly and aggregated to produce the forecast for time $t+1$.

In summary, the use of queries, keys, and values in multi-head attention enables the model to focus on different parts of the input sequence and learn complex temporal relationships, making it highly effective for sequence-based prediction tasks.

\subsection*{LSTM}
\label{lstm}

For comparison purposes, we decided to use a less complex Long Short Term Memory model (LSTM) which was suggested in many previous studies 
on time series forecasting. LSTM was firstly introduced in \cite{10.1162/neco.1997.9.8.1735}.

\begin{figure}[htbp]
  \centering
   \includegraphics[width=0.9\linewidth]{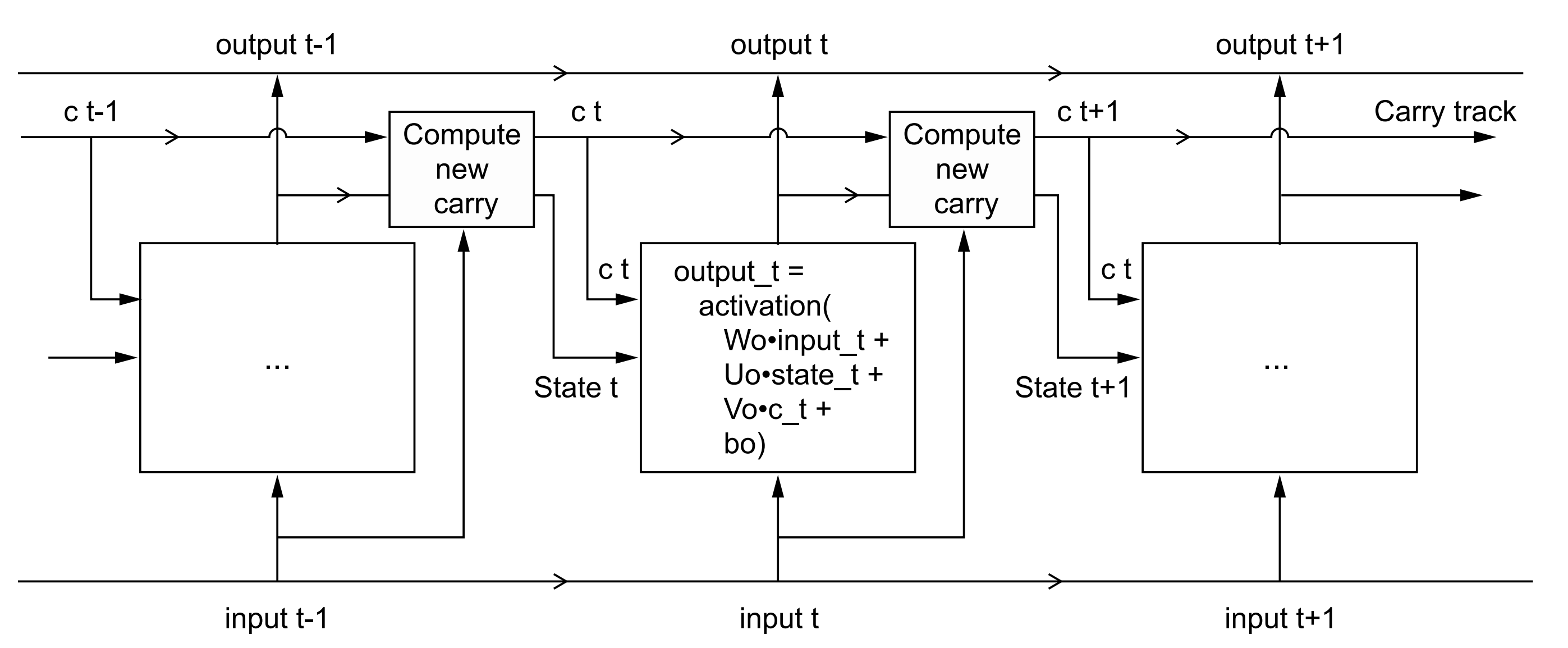}
   \caption{LSTM cells presented in this Fig. show the information flow between the main LSTM gates: input, output, and forget.  Source: \cite{chollet_2021}.}
   \label{fig:lstmanatomy}
\end{figure}

LSTM models process input step-by-step using memory cells and gating mechanisms to capture temporal dependencies. Their recurrent design makes them suitable for tasks where sequence order matters, like time-series prediction, speech recognition, and certain NLP applications. However, their sequential nature limits parallelization, leading to slower training on long sequences or large datasets. While LSTMs manage short- and mid-range dependencies effectively, they often fail to retain information across very long sequences. The architecture of an LSTM model is shown in Fig. \ref{fig:lstmanatomy}.

In contrast, Transformer models rely on a self-attention mechanism that allows them to capture dependencies across an entire sequence simultaneously, regardless of distance. This parallel processing capability speeds up training significantly and enables the model to scale well with large datasets, making Transformers ideal for tasks requiring long-range dependencies, such as machine translation and text summarization. Additionally, Transformers incorporate positional encoding to track the order of tokens without needing a recurrent structure. As a result, they excel in natural language processing and have been successfully adapted for applications in computer vision and other domains that benefit from highly scalable and efficient training.

\subsection{Model Hyperparameters}

Our transformer model consists of two multi-head attention layers and one single neuron dense layer on the output. The sequence length (key) is set to 3, and we use four parallel heads. Each of the LSTM layers uses tanh activation function (to retain negative values). L2 regularization (1\text{e-}6) and dropout (0.03) are also applied to each of these layers. The first two layers return sequences with the same shape as the input sequence (full sequence), and the last layer returns only the last output.

To train the model we used the Adam optimizer - a stochastic gradient descent optimizer with momentum (estimating first-order and second-order moments). The learning rate of the optimizer was set to 0.5. The summary of selected hyperparameters can be found in Table \ref{tab:Hyperparameters}.

\begin{table}[htbp]
\caption{Selected values of hyperparameters.}
\vspace{-10pt}
\centering
\begin{center}
\begin{tabular}[t]{ll}
\hline
  \textbf{Hyperparameter} & \textbf{Selected Value}\\
\hline
 No. hidden layers (LSTM/Tran.) & 3/2\\
 No. neurons (LSTM) & 512/256/128\\
 Activation function (LSTM) & \textit{tanh}\\
 Dropout rate (LSTM/Tran.) & 0/0.3\\
 l2 regularizer (LSTM/Tran.) & 1\text{e-}6/0.02 \\
 Optimizer & Adam\\
 Learning rate (LSTM/Tran.) & 0.5/0.01\\
 Train/test size & 252/252\\
 Batch size & max\\
 Sequence length & 4\\
 Num heads (Tran.) & 4\\
 Key (value) dim (Tran.) & 64\\
 No. attention layers (Tran.) & 2 \\
\hline
\end{tabular}
\end{center}
\scriptsize
\vspace{-5pt}
Note: Hyperparameters used in this study for the LSTM and Transformer model.
\label{tab:Hyperparameters}
\end{table}

\subsection{Data and Research description}

We use simple returns, based on daily data from 2004-01-02 to 2024-10-24 for S\&P500, XOM, and JPM, as representatives of the equity market and BTC, ETH, and LTC as representatives of cryptocurrency markets (starting at 2014-09-17 for BTC and ETH, and 2015-08-07 for ETH). The selection of such three equities was based on the desire to select one very representative equity index and two shares that have been part of this equity index for many years. In the case of cryptocurrency selection, we focused on cryptos with the highest market cap and the longest time series available.

Based on the presented methodology we were able to plan our research in the following way:

\begin{itemize}

\item For the training set, we used an expanding window approach, with the size of the first window set to 252/365 trading days (one year). The validation set was set size to 33\% of the training set. The test set size was also 252/365 days.
    
\item The input sequence size was set to 3.
    
\item We used the ReLU activation function on the last neuron to obtain only zero or positive values (for Long Only strategies)
    
\item The output of the model was a single number predicting the next return value.
    
\item Based on the sign of the predicted return value we assigned -1, 0, and 1 signals, depending on the strategy.

\item Two models used in this research are: 1) Long Short-Term Memory network (LSTM) which is a well-known type of deep recurrent network, 2) Transformer based neural network 

\item We use a rolling walk-forward procedure for training and testing, to avoid common drawbacks in this type of research

\item A custom loss function (MADL) was created as the network performance metric and was used during the training process.

\item Strategy performance metrics - equity line and strategy-specific performance metrics (aRC, aSD, MD, MLD, IR, IR\*, IR\*\*, nObs).

\end{itemize}

\subsection{Model Training}

For training and prediction, we used a walk-forward validation/expanding window approach. In the first iteration, the model was trained on one year of data (equal to the train set length) and then used for predictions over the next year (equal to the test set length). After that, the window was expanded by another year of data (up to 4 years) and the model was retrained. A single return value was predicted each time, based on the last 3 (sequence length) values.

A single iteration was trained for 300 epochs for LSTM and 50 epochs for the transformer. The model checkpoint callback function was used to store the best weights (parameters) of the model based on the lowest loss function value in a specific epoch. The weights were then used for prediction.

\subsection{Loss Function}

We use the loss function proposed by \cite{s22030917} and additionally developed and tested in \cite{MICHANKOW2024102375} which was built to improve the forecasting ability of ML models in algorithmic investment strategies (AIS).

\vspace{-2mm}
\begin{equation}
\textrm{MADL} =\frac{1}{N} \sum_{i=1}^{N} (-1) \times \textrm{sign}({R_{i} \times \hat{R}_{i}}) \times  \textrm{abs}(R_{i} )
\end{equation}
\label{fig:MADL}
\vspace{-5mm}

where: 
\begin{itemize}
    \item MADL is the Mean Absolute Directional Loss function, 
    \item \(R_{i}\) is the observed return on interval \(i\), 
    \item \(\hat{R}_{i}\) is the predicted return on interval \(i\), 
    \item $\textrm{sign}(X)$ is the function which gives the sign of \(X\),
    \item $\textrm{abs}(X)$ is the function which gives the absolute value of \(X\) 
    \item \(N\) is the number of forecasts. 
\end{itemize}

When we approach the problem from this perspective, the function's value corresponds to the actual investment return given the forecasted direction. This allows the model to determine not only whether its prediction will yield a profit or loss, but also to estimate the magnitude of that expected gain or loss. MADL was purposefully developed for use with algorithmic investment strategies (AIS) beyond simply validating point predictions. Since our model minimizes this function, negative values indicate profitable strategy outcomes, while positive values signal potential losses.

\subsection{Performance Metrics}

Based on \cite{SlepaczukSakowskiZakrzewski2018} or \cite{kijewski2024predicting} the following performance metrics were calculated:

\begin{itemize}
    \item Annualized return compounded (aRC):

    \vspace{-15pt}
    \begin{align}
      \textrm{aRC} = \prod_{i = 1}^{n}{(r_i+1)}^{252/n} - 1
    \end{align}
    \vspace{-15pt}
    
    where:  
    $r_i$ - is the daily percentage return at time $i$  
    $n$ - is the number of trading days

    \item Annualized standard deviation (aSD):
    
    \vspace{-20pt}
    \begin{align}
      \textrm{aSD} = \sqrt{252} *  \frac{1}{n-1} * \sum_{i = 1}^{n}{(r_i - \bar{r})^2}
    \end{align}
    \vspace{-20pt}
    
    where $\bar{r}$ is the average daily percentage return 

    \item Maximum drawdown (MD):

        \vspace{-20pt}
        \begin{align}
        \textrm{MD} = sup_{x,y\ \epsilon\ \{[t_1,t_2]^2\ :\ x \leq y \}} \frac{P_x - P_y}{P_x}
        \end{align}
        \vspace{-20pt}
        
        where $P_t$ is the equity line level at time $t$

\end{itemize}

    \begin{itemize}
        
        \item Maximum Loss Duration (MLD): the longest time needed to surpass a maximum value ($m$) of the strategy returns. It is measured in years.

        \vspace{-20pt}
        \begin{align}
        \textit{MLD} = \max \frac{m_j - m_i}{N}
        \end{align}
        \vspace{-20pt}

        \item Information ratio* (IR*):
    
        \vspace{-20pt}
        \begin{align}
        IR^{*} = \frac{ARC}{aSD}
        \end{align}
        \vspace{-20pt}

        \item Information ratio** (IR**) - \textbf{we regard this metric as the most important in the evaluation of our final results:}

        \vspace{-20pt}
        \begin{align}
        IR^{**} = \frac{ARC*ARC*sign(ARC)}{aSD*MD}
        \end{align}
        \vspace{-20pt}

        \item Information ratio** (IR***)

        \vspace{-20pt}
        \begin{align}
        IR^{***} = \frac{ARC*ARC*ARC}{aSD*MD*MLD}
        \end{align}
        \vspace{-20pt}
        
    \end{itemize}

\subsection{Hardware and computation time}

The results for the tested models were obtained using R 4.3.1 along with Python 3.7.10. Deep learning libraries used for designing, training, and testing the network are Keras 2.13.0 and TensorFlow 2.13.0. Computer specification: AMD Ryzen 7 3700X 3,6GHz, 16GB RAM, NVIDIA GeForce RTX 2060 Super with 270 tensor cores. One full training (number of iterations $\times$ 50 epochs) lasted around 15 minutes.

\section{Results}
    \label{Results}

Based on the Research and Methodology description provided in Section \ref{Methodology and Data} we prepared the results that should be analyzed in two separate sets, the first one for equities and the second one for cryptocurrencies.

\begin{figure}[htpb]
\includegraphics[width=0.48\textwidth]{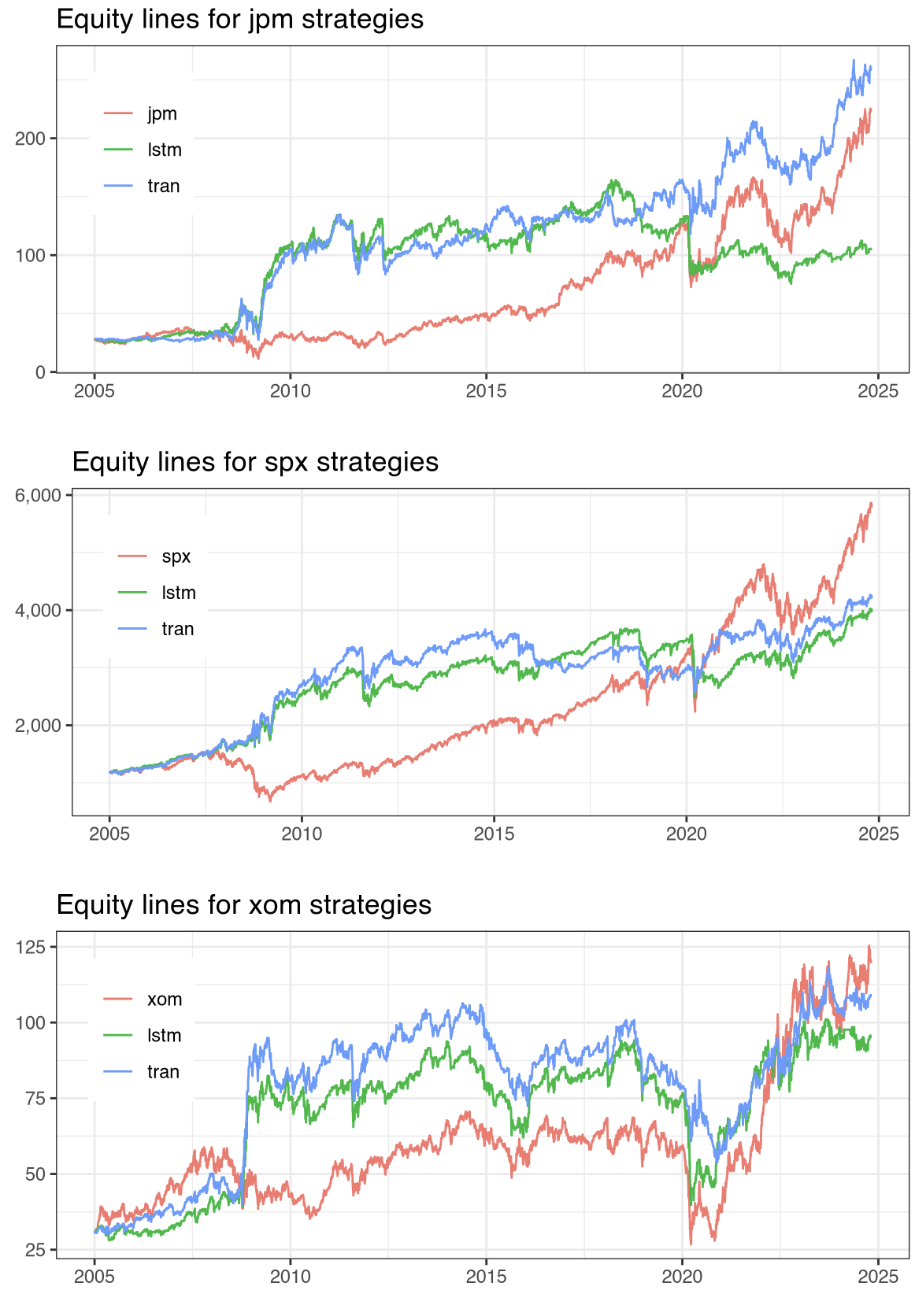}
\includegraphics[width=0.49\textwidth]{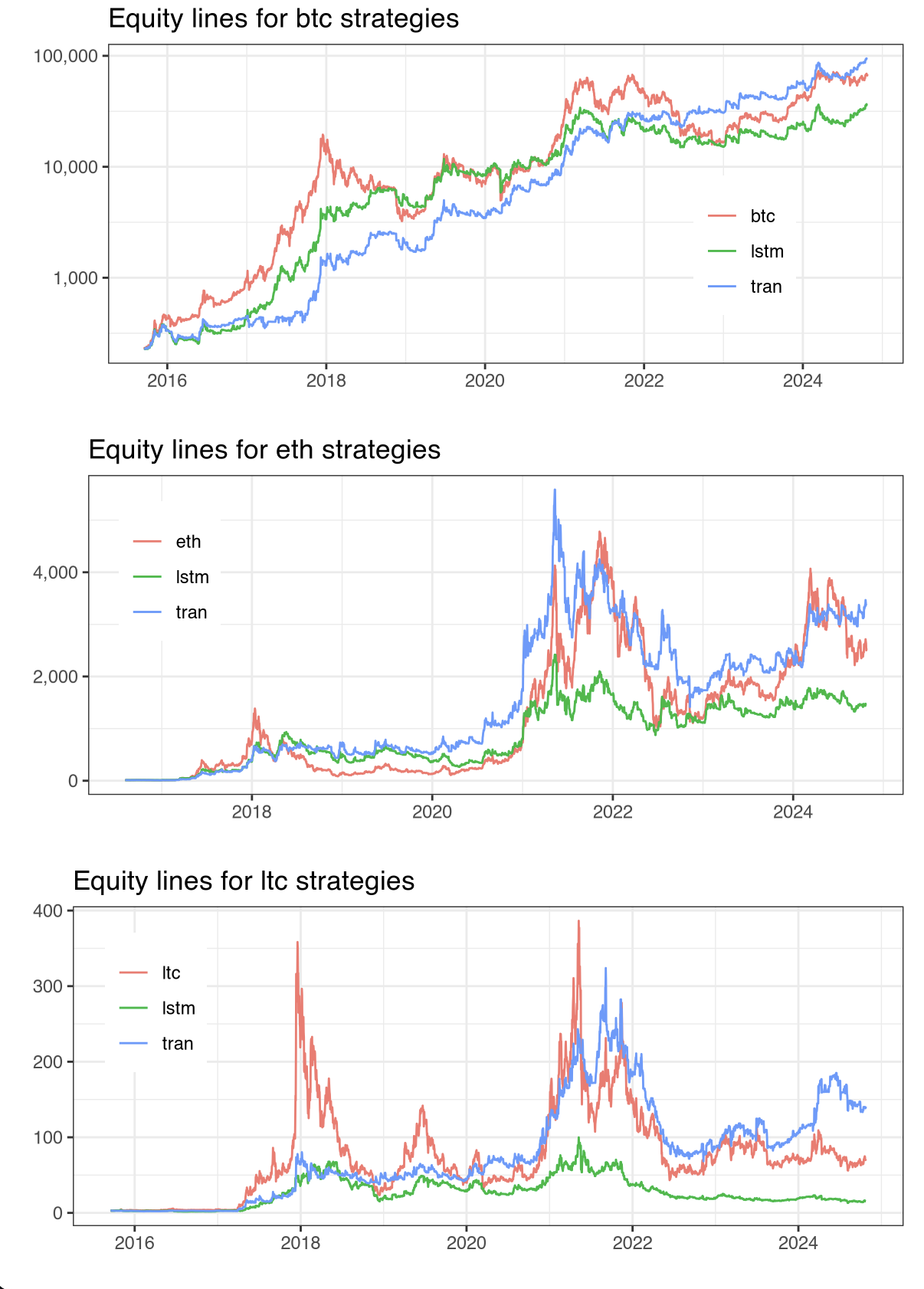}
\scriptsize 
\justifying

Note: Equity lines present the fluctuations of investment strategies for JPM (upper left panel), SPX (middle left panel), XOM (lower left panel), BTC (upper right panel), ETH (middle right panel), and LTC (lower right panel) for strategies based on LSTM and transformer with Mean Absolute Directional Loss function in the period between Jan 3, 2005 and Oct 24, 2024 (JPM, SPX, XOM) and Sep 17, 2015 (BTC, LTC), Aug 6, 2016 (ETH), and Oct 24, 2024 (BTC, ETH, LTC). Additionally, the buy\&hold (B\&H) strategies were included as a benchmarks.\caption{Equity lines for JPM, SPX, XOM, BTC, ETH and LTC.}

\label{fig:eql_equity}
\end{figure}

Table \ref{tab:results_equity} presents the results for three equities (S\&P500 index, Exxon Mobil Corp, and JPMorgan) showing that in the case of each analyzed time series risk-adjusted return metrics (IR*, IR**, and IR***) for Transformer models are higher in comparison to LSTM models and Buy\&Hold strategy. Moreover, equity curves described in left panel of Fig. \ref{fig:eql_equity} confirm the superior performance of Transformer models. 

\begin{table}[hbp]
\centering\fontsize{8}{8}\selectfont
\caption{Performance measures for SPX, JPM, and XOM}
\label{tab:results_equity}

\begin{tabular}{lrrrrrrrrr}
    \toprule
    Model & aRC & aSD & MD & MLD & IR* & IR** & IR*** & nObs & nTrades\\
    \midrule
\\
\multicolumn{10}{l}{\textbf{JPM}}
\vspace{3pt}
\\
B\&H & 11.06 & 36.42 & 70.12 & 5.82 & 0.30 & 0.048 & 0.001 & 4987 & 2\\
LSTM & 6.91 & 27.53 & 54.01 & 6.62 & 0.25 & 0.032 & 0.000 & 4987 & 1378\\
TRANS & 11.89 & 26.89 & 56.01 & 4.00 & \textbf{0.44} & \textbf{0.094} & \textbf{0.003} & 4987 & 1672
\vspace{3pt}
\\
\hline
\\
\multicolumn{10}{l}{\textbf{SPX}}
\vspace{3pt}
\\
B\&H & 8.29 & 19.22 & 56.78 & 5.46 & 0.43 & 0.063 & 0.001 & 4987 & 2\\
LSTM & 6.25 & 14.64 & 32.42 & 5.67 & 0.43 & 0.082 & 0.001 & 4987 & 1594\\
TRANS & 6.56 & 14.05 & 30.04 & 7.01 & \textbf{0.47} & \textbf{0.102} & 0.001 & 4987 & 1698
\vspace{3pt}
\\
\hline
\\
\multicolumn{10}{l}{\textbf{XOM}}
\vspace{3pt}
\\
B\&H & 7.06 & 26.67 & 62.11 & 7.58 & 0.26 & 0.030 & 0.000 & 4987 & 2\\
LSTM & 5.86 & 19.41 & 57.78 & 4.43 & 0.30 & 0.031 & 0.000 & 4987 & 1290\\
TRANS & 6.56 & 18.89 & 49.35 & 8.66 & \textbf{0.35} & \textbf{0.046} & 0.000 & 4987 & 1723
\vspace{3pt}
\\

  \bottomrule
\end{tabular}

\scriptsize \justifying
Note: aRC - annualized return compounded, aSD - annualized standard deviation, MD - Maximum Drawdown, IR*, IR**, IR*** - Information Ratio and its two modifications, MLD - Maximum Loss Duration, the longest time needed to surpass a maximum value of the strategy returns, measured in years,  nObs - the number of observations, nTrades - the number of trades, which is the number of all changes in position on the analyzed asset. \textit{B\&H} stands for Buy\&Hold strategy results. \textit{LSTM} indicates for LSTM strategy results. \textit{TRANS} stands for Transformer strategy results. 
\end{table}

Similar conclusions can be drawn from Table \ref{tab:results_crypto}. Once again we can see that the most efficient results can be obtained for Transformer models in the case of every cryptocurrency. Right panel of Figure \ref{fig:eql_equity} showing equity curves confirms the results from Table \ref{tab:results_crypto}.

\begin{table}[h]
\centering\fontsize{8}{8}\selectfont
\caption{Performance measures for BTC, EHT and LTC}
\label{tab:results_crypto}

\begin{tabular}{lrrrrrrrrr}
    \toprule
    Model & aRC & aSD & MD & MLD & IR* & IR** & IR*** & nObs & nTrades\\
    \midrule
\\
\multicolumn{10}{l}{\textbf{BTC}}
\vspace{3pt}
\\
B\&H & 86.35 & 69.49 & 83.40 & 2.96 & 1.24 & 1.287 & 0.376 & 3328 & 2\\
LSTM & 73.61 & 49.96 & 55.93 & 2.98 & 1.47 & 1.939 & 0.480 & 3328 & 1254\\
TRANS & 92.86 & 47.12 & 34.53 & 0.78 & \textbf{1.97} & \textbf{5.301} & \textbf{0.632} & 3328 & 1130
\vspace{3pt}
\\
\hline
\\
\multicolumn{10}{l}{\textbf{ETH}}
\vspace{3pt}
\\
B\&H & 93.51 & 92.41 & 93.91 & 3.02 & 1.01 & 1.008 & 0.312 & 3005 & 2\\
LSTM & 80.65 & 64.22 & 71.62 & 3.47 & 1.26 & 1.414 & 0.329 & 3005 & 1557\\
TRANS & 100.47 & 66.84 & 74.66 & 3.47 & \textbf{1.50} & \textbf{2.022} & \textbf{0.586} & 3005 & 1031
\vspace{3pt}
\\
\hline
\\
\multicolumn{10}{l}{\textbf{LTC}}
\vspace{3pt}
\\
B\&H & 28.98 & 85.48 & 93.45 & 4.78 & 0.34 & 0.105 & 0.006 & 3329 & 2\\
LSTM & 14.45 & 58.93 & 86.84 & 4.78 & 0.25 & 0.041 & 0.001 & 3329 & 1348\\
TRANS & 36.55 & 62.87 & 78.92 & 4.33 & \textbf{0.58} & \textbf{0.269} & \textbf{0.023} & 3329 & 1210
\vspace{3pt}
\\
  \bottomrule
\end{tabular}

\scriptsize \justifying

Note: Note: aRC - annualized return compounded, aSD - annualized standard deviation, MD - Maximum Drawdown, IR*, IR**, IR*** - Information Ratio and its two modifications, MLD - Maximum Loss Duration, the longest time needed to surpass a maximum value of the strategy returns, measured in years,  nObs - the number of observations,  nTrades - the number of trades, which is the number of all changes in position on the analyzed asset. \textit{B\&H} stands for Buy\&Hold strategy results. \textit{LSTM} indicates for LSTM strategy results. \textit{TRANS} stands for Transformer strategy results. 
\end{table}

The presented results confirm our initial presumptions that a more sophisticated and complex model, like a Transformer, used with proper Loss function can enable us to construct efficient investment strategies.

\section{Conclusions}
\label{Conclusions}

In this study, we evaluate the application of the Mean Absolute Directional Loss function (\cite{s22030917}) in algorithmic trading with two machine learning algorithms: the transformer model (\cite{vaswani2023attentionneed}) and the LSTM (\cite{10.1162/neco.1997.9.8.1735}). The models were applied to daily data from six assets (cryptocurrencies, including Bitcoin, Ethereum, and Litecoin, and equity stocks, including JP Morgan, S\&P 500, and Exxon Mobil). The walk-forward procedure was used to include the out-of-sample period, which extended over eight years. 

The results show that we successfully adapted the basic transformer architecture to produce trading strategies yielding abnormal risk-adjusted returns. The transformer outperforms the Buy\&Hold and LSTM-based strategies for both types of assets under investigation. This architecture produces higher risk-adjusted returns compared with both the LSTM and the Buy\&Hold strategy.

Our contribution to the literature is threefold. First, we demonstrate the application of an appropriate loss function (MADL) within machine learning models to generate trading signals. Second, we assess the advantages of using transformer models over LSTM models in algorithmic trading. Third, we apply a rigorous methodology across six assets, carefully controlling for overfitting, implementing a walk-forward procedure, and extending the out-of-sample period to over nine years for equities and over eight years for cryptocurrency assets.

The findings from this study carry several potential policy implications, particularly for financial market regulation and algorithmic trading oversight. First, the demonstrated ability of transformer models to consistently outperform traditional strategies highlights the growing role of advanced machine learning in generating high risk-adjusted returns. This may prompt regulatory bodies to consider new guidelines for algorithmic trading practices, especially regarding transparency and risk management. Furthermore, given the long out-of-sample testing period and robust methodology employed, these findings may encourage policy discussions around implementing stricter standards for the validation and monitoring of algorithmic models to safeguard against overfitting and ensure consistent performance. Finally, as these advanced models could widen the gap between retail and institutional investors, policies may be required to promote equitable access to AI-driven trading technologies.

Further research should concentrate on extensive sensitivity analysis, including a wide range of hyperparameters included in tuning phases, using extended datasets in terms of higher frequency and even longer out-of-time periods. It would be beneficial to verify the application of the MADL function in other types of deep networks and machine learning models. Finally, the MADL function could be still improved to address the problem of its non-differentiability in certain areas (\cite{s22030917}).

\bibliography{main}{}

\end{document}